\documentclass[letterpaper,prl,twocolumn,showpacs,superscriptaddress]{revtex4} 
\usepackage{graphicx}
\usepackage{times}
\newcommand{\expec}[1]{\langle#1\rangle}  

\begin{document}
   
\title{Polarization Squeezing of Continuous Variable Stokes Parameters}

\author{Warwick P. Bowen}
\affiliation{Department of Physics, Faculty of Science, Australian
National University, ACT 0200, Australia}

\author{Roman Schnabel} 
\affiliation{Department of Physics, Faculty of Science, Australian
National University, ACT 0200, Australia}

\author{Hans -A. Bachor} 
\affiliation{Department of Physics, Faculty of Science, Australian
National University, ACT 0200, Australia}

\author{Ping Koy Lam} 
\affiliation{Department of Physics, Faculty of Science, Australian
National University, ACT 0200, Australia}

\begin{abstract}
We report the first direct experimental characterization of continuous
variable quantum Stokes parameters.  We generate a continuous wave
light beam with more than 3~dB of simultaneous squeezing in three of
the four Stokes parameters.  The polarization squeezed beam is
produced by mixing two quadrature squeezed beams on a polarizing beam
splitter.  Depending on the squeezed quadrature of these two beams the
quantum uncertainty volume on the Poincar\'{e} sphere became a `cigar'
or `pancake'-like ellipsoid.
\end{abstract}


\maketitle

The quantum properties of the polarization of light have received much
attention in the single photon regime where fundamental problems of
quantum mechanics, related to Bell's inequality and the
Einstein-Podolski-Rosen (EPR) paradox, have been examined
\cite{AGR82}.  In comparison quantum polarization states in the
continuous variable regime have received little attention.  Grangier
et al.  \cite{GSYL87} generated a polarization squeezed beam using an
optical parametric process to improve the sensitivity of a
polarization interferometer.  Other schemes using Kerr-like media and
optical solitons in fibers have also been proposed \cite{AAC98,
KLLRS01} but have not yet been produced experimentally.  This paper
presents the generation of a new polarization squeezed state.  A
stably locked beam with better than 3~dB of squeezing in three Stokes
parameters simultaneously was produced utilizing two bright quadrature
squeezed beams.  We present the first direct characterization of the
polarization quantum noise on a continuous wave light beam allowing an
experimental observation of polarization commutation relations in the
continuous variable regime.

Both the production of new continuous variable quantum polarization
states and the ability to accurately characterize them are necessary
for these states to fulfill their potential in the field of quantum
information.  They can be carried by a bright laser beam providing
high bandwidth capabilities and therefore faster signal transfer rates
than single photon systems.  Perhaps surprisingly they retain the
single photon advantage of not requiring the universal local
oscillator necessary for other proposed continuous variable quantum
networks.  Mapping of quantum states from photonic to atomic media is
a crucial element in most proposed quantum information networks. 
Continuous variable polarization states are the only continuous
variable state for which this mapping has been experimentally
demonstrated \cite{HSSP99}.

The polarization state of a light beam in classical optics can be
visualized as a Stokes vector on a Poincar\'{e} sphere (Fig.~1) and is
determined by the four Stokes parameters \cite{Sto52}: $S_{0}$
represents the beam intensity whereas $S_{1}$, $S_{2}$, and $S_{3}$
characterize its polarization and form a cartesian axes system.  If
the Stokes vector points in the direction of $S_{1}$, $S_{2}$, or
$S_{3}$ the polarized part of the beam is horizontally, linearly at
45$^\circ$, or right-circularly polarized, respectively.  Two beams
are said to be opposite in polarization and do not interfere if their
Stokes vectors point in opposite directions.  The quantity $\,S =
(S_{1}^{2}+S_{2}^{2}+S_{3}^{2})^{1/2}\,$ is the radius of the
classical Poincar\'{e} sphere and the fraction $\,S / S_{0}\,$
($\,0\!<\!S/S_{0}\!<\!1\,$) is called the degree of polarization.  For
quasi-monochromatic laser light which is almost completely polarized
$S_{0}$ is a redundant parameter, completely determined by the other
three parameters ($S_{0}\!=\!S$ in classical optics).  All four Stokes
parameters are accessible from the simple experiments shown in
Fig.~2.\\
%
\begin{figure}[]
  \centerline{\includegraphics[width=8.5cm]{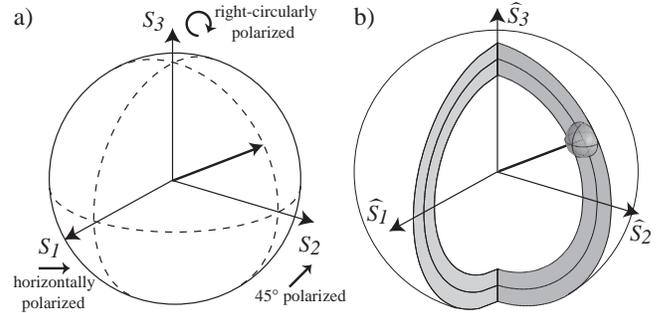}}
  \caption{Diagram of a) classical and b) quantum Stokes vectors
  mapped on a Poincar\'{e} sphere; the ball at the end of the quantum
  vector visualizes the quantum noise in $\rm \hat S_{1}$, $\rm \hat
  S_{2}$, and $\rm \hat S_{3}$; and the non-zero quantum sphere
  thickness visualizes the quantum noise in $\rm \hat S_{0}$.}
    \label{Poincare}
\end{figure}
%
%
\begin{figure}[]
  \centerline{\includegraphics[width=8.5cm]{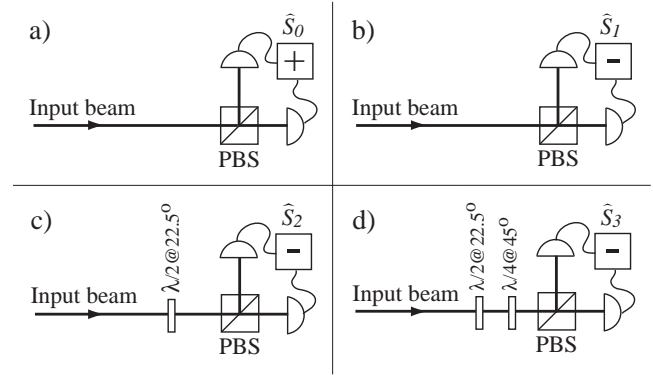}}
  \vspace{-3mm}
  \caption{Apparatus required to measure each of the Stokes parameters.
  PBS: polarizing beam splitter, $\lambda/2$ and $\lambda/4$:
  half- and quarter-wave plates respectively, the plus and
  minus signs imply that an electrical sum or difference has been
  taken.}
    \label{MeasuringStokes}
\end{figure}
The Stokes parameters are observables and therefore can be associated
with Hermitian operators. Following \cite{JauchRohrlich76} and
\cite{Robson74} we define the quantum-mechanical analogue of the
classical Stokes parameters for pure states in the commonly used notation:
\begin{eqnarray}
\label{stokes}
\hat S_{0}& \!= \hat a_{H}^{\dagger} \hat a^{ }_{H} + \hat a_{V}^{\dagger}
\hat a_{V} ~,\hspace{2mm} 
\hat S_{2}& \!=
\hat a_{H}^{\dagger} \hat a_{V}^{ } e^{i\theta} \!+ \hat a_{V}^{\dagger}
\hat a_{H}^{ } e^{-i\theta} ,\;\;\;\;\;\;\\\nonumber 
\hat S_{1}& \!= \hat a_{H}^{\dagger} \hat
a_{H}^{ } - \hat a_{V}^{\dagger} \hat a_{V}^{ } ~,\hspace{2mm} 
\hat S_{3}& \!=i\hat
a_{V}^{\dagger} \hat a_{H}^{ } e^{-i\theta} \!-i\hat a_{H}^{\dagger} \hat
a_{V}^{ } e^{i\theta} ,\;\;\;\;\;\;
\end{eqnarray}
where the subscripts $H$ and $V$ label the horizontal and vertical
polarization modes respectively, $\theta$ is the phase shift between
these modes, and the $\hat a^{ }_{H,V}$ and $\hat a_{H,V}^{\dagger}$
are annihilation and creation operators for the electro-magnetic field
in frequency space.  The commutation relations of these operators
%
\begin{equation}
[\hat a_{k}, \hat a_{l}] = \delta_{kl} 
~,\hspace{4mm}\mbox{with}\hspace{4mm} k,l \in\{H,V\} ~,
\label{acomrel}
\end{equation}
directly result in Stokes operator commutation relations,
\begin{equation}
[\hat S_{1}, \hat S_{2}] =  2 i \hat S_{3} ~,\hspace{2mm}
[\hat S_{2}, \hat S_{3}] =  2 i \hat S_{1} ~,\hspace{2mm}
[\hat S_{3}, \hat S_{1}] =  2 i \hat S_{2} ~\,.
\label{Scomrel}
\end{equation}
%
Apart from a normalization factor, these relations are identical to
the commutation relations of the Pauli spin matrices.  In fact the
three Stokes parameters in Eq.  (\ref{Scomrel}) and the three Pauli
spin matrices both generate the special unitary group of symmetry
transformations SU(2) of Lie algebra \cite{Kaku93}.  Since this group
obeys the same algebra as the three-dimensional rotation group,
distances in three dimensions are invariant.  Accordingly the operator
$\hat S_{0}$ is also invariant and commutes with the other three
Stokes operators ($\,[\hat S_{0}, \hat S_{j}] = 0,
\;\mbox{with}\;j=1,2,3$).  It can be shown from Eqs.~(\ref{stokes})
and (\ref{acomrel}) that the quantum Poincar\'{e} sphere radius is
different from its classical analogue, $\,\expec{\hat S} = \expec{\hat
S_{0}^{2} + 2 \hat S_{0}}^{1/2}\,$.
The noncommutability of the Stokes operators $\hat S_{1}$, $\hat S_{2}$ and
$\hat S_{3}$ precludes the simultaneous exact measurement of their physical
quantities. Their mean values $\,\expec{\hat S_{j}}\,$ and variances
$\,V_j=\expec{\hat S_{j}^2}\!-\!\expec{\hat S_{j}}^2\,$ are restricted by the
uncertainty relations~\cite{JauchRohrlich76}
\begin{equation}
V_1 V_2 \ge |\expec{\hat S_{3}}|^2 ~, ~ V_2 V_3 \ge |\expec{\hat
S_{1}}|^2 ~, ~ V_3 V_1 \ge |\expec{\hat S_{2}}|^2 ~.
\label{uncer}
\end{equation}
In general this results in non-zero variances in the individual Stokes
parameters as well as in the radius of the Poincar\'{e} sphere (see
Fig.~\ref{Poincare}b)).  Recently it has been shown that these
variances may be obtained from the frequency spectrum of the
electrical output currents of the setups shown in
Fig.~\ref{MeasuringStokes}~\cite{KLLRS01}.

It is useful to express the Stokes operators of Eq.~(\ref{stokes}) in
terms of quadrature amplitudes of the horizontally and vertically
polarized components of the beam.  The creation and annihilation
operators can be expressed as sums of real classical amplitudes
$\alpha^{ }_{H,V}$ and quadrature quantum noise operators $\delta \hat
X_{H,V}^{+}$ and $\delta \hat X_{H,V}^{-}$ \cite{WallsMilburn}
\vspace{-1mm}
\begin{equation}
\hat a^{ }_{H,V} = \alpha^{ }_{H,V} + \frac{1}{2}(\delta\hat X_{H,V}^+ + i\delta\hat
X_{H,V}^-)  ~.
\label{lin}
\end{equation}
If the variances of the noise operators are much smaller than
the coherent amplitudes then, to first order in the noise operators, the Stokes
operator mean values are 
\begin{eqnarray}
\label{expec}
\expec{\hat S_{0}}&=& \alpha_{H}^2+\alpha_{V}^2 = \expec{\hat 
n}~,\hspace{2mm}
\expec{\hat S_{2}} = 2 \alpha^{ }_{H} \alpha^{ }_{V} \,\mbox{cos}\theta~,\\\nonumber
\expec{\hat S_{1}}&=&\alpha_{H}^2-\alpha_{V}^2                  
~,\hspace{11.5mm}
\expec{\hat S_{3}} = 2 \alpha^{ }_{H} \alpha^{ }_{V} \,\mbox{sin}\theta       ~,
\end{eqnarray}
where $\,\expec{\hat n}\,$ is the expectation value of the photon
number operator.  For a coherent beam the expectation value and
variance of $\hat n$ have the same magnitude, this magnitude equals
the conventional shot-noise level.  The variances of the Stokes
parameters are given by
\begin{eqnarray}
\label{var}
V_0=V_1=& \!\alpha_{H}^2 \expec{(\delta\hat X_{H}^+)^2}\! + 
\!\alpha_{V}^2 \expec{(\delta\hat X_{V}^+)^2},\;\;\;\;\nonumber\\
%
%
V_2(\theta\!=\! 0)\! = \! V_3(\theta \! = \! \pi \!  /2) = & 
\!  \alpha_{V}^2 \expec{(\delta\hat X_{H}^+)^2} \!  + \!  \alpha_{H}^2 
\expec{(\delta \hat X_{V}^+)^2},\;\;\;\;\\
V_3(\theta\!=\!0)\!= \!V_2(\theta\!=\!\pi\!/2)=& \!\alpha_{V}^2 
\expec{(\delta\hat X_{H}^-)^2}\! + \!\alpha_{H}^2 \expec{(\delta\hat 
X_{V}^-)^2},\;\;\;\;\nonumber
\nonumber
\end{eqnarray}
The variances of the noise operators in the above equation are
normalized to one for a coherent beam.  Therefore the variances of the
Stokes parameters of a coherent beam are all equal to the shot-noise
of the beam.  For this reason a Stokes parameter is said to be
squeezed if its variance falls below the shot-noise of an equal power
coherent beam.  Although the decomposition to the $H,V$-polarization
axis of Eqs.~(\ref{var}) is independent of the actual procedure of
generating a polarization squeezed beam, it becomes clear that two
overlapped quadrature squeezed beams can produce a single polarization
squeezed beam.  We choose the two specific angles of $\,\theta \!=\! 
0$ and $\pi/2\,$ in Eqs.~(\ref{var}) corresponding to our actual
experimental setup.  If the polarization squeezed beam is generated
from two amplitude squeezed beams $\hat S_0$ and two additional Stokes
parameters can in theory be perfectly squeezed while the fourth is
anti-squeezed.  In this case the uncertainty volume formed by the
noise variances of the Stokes parameters in the Poincar\'{e} space
becomes a ``cigar''-like ellipsoid.  If the polarization squeezed beam
is generated from two phase squeezed beams the uncertainty volume
becomes a ``pancake''-like ellipsoid (see Fig.~\ref{Prettyresults}).
\begin{figure}[b]
 \centerline{\includegraphics[width=8.5cm]{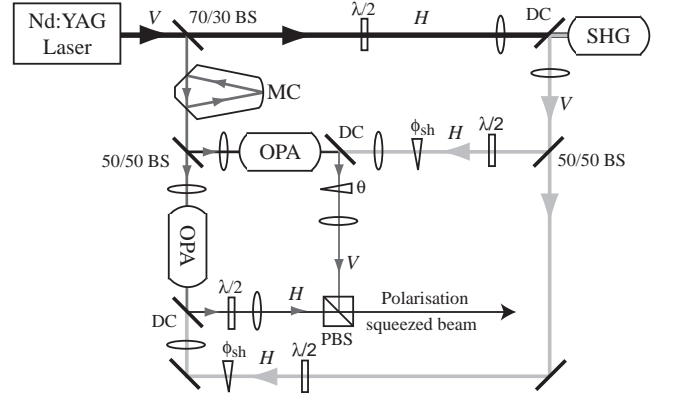}}
\caption{Schematic of the polarization squeezing experiment.  BS: beam 
splitter, DC: dichroic beam splitter, $\lambda/2$: half-wave plate, 
$\phi_{sh}$: phase shift between 532~nm and 1064~nm light at the OPAs, 
$\theta$: phase shift between quadrature squeezed beams, PBS: 
polarizing beam splitter.}
   \label{experiment}
\end{figure}

The experimental setup used to generate the polarization squeezed
beams is shown in Fig.~\ref{experiment}.  Two quadrature squeezed
beams were produced in a pair of spatially separated optical
parametric amplifiers (OPAs).  Past experiments requiring two
quadrature squeezed beams commonly used a single ring resonator with
two outputs \cite{Ou}; with two independent OPAs the necessary
intracavity pump power is halved, this reduces the degradation of
squeezing due to green-induced infrared absorption (GRIIRA)
\cite{Furukawa}.  The OPAs were optical resonators constructed from
hemilithic MgO:LiNbO$_{3}$ crystals and output couplers.  The
reflectivities of the output couplers were 96\% and 6\% for the
fundamental (1064~nm) and the second harmonic (532~nm) laser modes,
respectively.  The OPAs were pumped with single-mode 532~nm light
generated by a 1.5~W Nd:YAG non-planar ring laser and frequency
doubled in a second harmonic generator (SHG).  The SHG was of
identical structure to the OPAs but with 92\% reflectivity at 1064~nm. 
Each OPA was seeded with 1064~nm light after spectral filtering in a
modecleaner.  This seed enabled control of the length of the OPAs. 
The coherent amplitude of each OPAs output was a deamplified/amplified
version of the seed coherent amplitude; the level of amplification was
dependent on the phase difference between pump and seed ($\phi_{sh}$). 
Locking to deamplification or amplification provided amplitude or
phase squeezed outputs, respectively.

The two quadrature squeezed beams were combined with orthogonal
polarization on a polarizing beam splitter, similarly to a scheme
proposed for solitons \cite{KLLRS01}.  This produced an output beam
with Stokes parameter variances as given by Eqs.~(\ref{var}).  Both
input beams had equal power ($\alpha_{H}^{ } \!  = \!  \alpha_{V}^{ } \!=\! 
\alpha/\sqrt{2}$) and both were squeezed in the same quadrature.  The
Stokes parameters and their variances were determined as shown in
Fig.~\ref{MeasuringStokes}.  The beam was split on a polarizing beam
splitter and the two outputs were detected on a pair of high quantum
efficiency photodiodes with 30~MHz bandwidth; the resulting
photocurrents were added and subtracted to yield the means and
variances of $\hat S_{0}$ and $\hat S_{1}$.  To measure $\expec{\hat
S_{2}}$ and $V_2$ the polarization of the beam was rotated by
45$^\circ$ with a half-wave plate before the polarizing beam splitter
and the detected photocurrents were subtracted.  To measure
$\expec{\hat S_{3}}$ and $V_3$ the polarization of the beam was
rotated by 45$^\circ$ with a half-wave plate and a quarter-wave plate
was introduced before the polarizing beam splitter such that a
horizontally polarized input beam became right-circular.  Again the
detected photocurrents were subtracted.  The relative phase between
the quadrature squeezed input beams $\theta$ was locked to $\pi/2$
rads producing in all cases a right-circularly polarized beam with
Stokes parameter means of $\langle \hat S_{1} \rangle \!=\!  \langle
\hat S_{2} \rangle \!=\!  0$ and $\langle \hat S_{0} \rangle \!=\! 
\langle \hat S_{3} \rangle \!=\!  |\alpha|^{2}$.  There is no
fundamental bias in the orientation of the quantum Stokes vector. By
varying the angle of an additional half-wave plate in the polarization
squeezed beam and by varying $\theta$ any orientation may be achieved. 
In fact the experiment discussed here was also carried out with
$\theta$ locked to 0 rads, in this case the Stokes vector was rotated
to point along $\hat S_{2}$ and the quantum noise was similarly
rotated.  Near identical results were obtained but on alternative
Stokes parameters.

Each quadrature squeezed beam had an overall detection efficiency of
73\%.  The loss came primarily from four sources: loss in escape from
the OPAs (14\%), detector inefficiency (7\%), loss in optics (5\%),
and mode overlap mismatch between the beams (4\%).  Depolarizing
effects are thought to be another significant source of loss for
some polarization squeezing proposals \cite{APu89}.  In our scheme the
non-linear processes (OPAs) are divorced from the polarization
manipulation (wave plates and polarizing beam splitters), and
depolarizing effects are insignificant.

As discussed earlier the Stokes parameters of a beam are squeezed if
their variances fall below the shot-noise level of an equally intense
coherent beam.  This shot-noise level was used to calibrate the
measurements presented here, it was determined by operating a single
OPA without the second harmonic pump.  The seed power was adjusted so
that the output power was equal to that of the polarization squeezed
beam.  In this configuration the detection setup for $\hat S_{2}$ (see
Fig.~\ref{MeasuringStokes}c)) functions exactly as a homodyne detector
measuring vacuum noise scaled by the OPA output power the variance of
which is the shot-noise.  Throughout each experimental run the sum of
output powers from the two OPAs was monitored and was always within
2\% of the power of the coherent calibration beam.  This led to a
conservative error in our frequency spectra of $\pm$0.04~dB.
\begin{figure}[t]
 \centerline{\includegraphics[width=8.5cm]{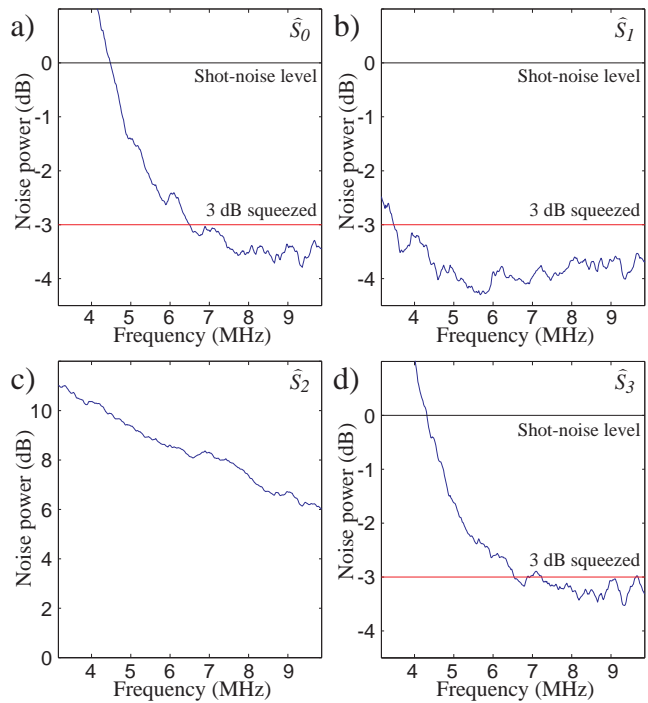}}
 \vspace{-2mm}
 \caption{Measured variance spectra of quantum noise on a) $\hat
 S_{0}$, b) $\hat S_{1}$, c) $\hat S_{2}$, and d) $\hat S_{3}$ for
 locked amplitude squeezed input beams; normalized to shot-noise.}
 \label{Ampresults}
\end{figure}

The shot-noise and Stokes parameter variances presented here were
taken with a Hewlett-Packard E4405B spectrum analyzer over the range
from 3 to 10~MHz.  The darknoise of the detection apparatus was always
more than 4~dB below the measured traces and was taken into account. 
Each displayed trace is the average of three measurement results
normalized to the shot-noise and smoothed over the resolution
bandwidth of the spectrum analyzer which was set to 300~kHz.  The
video bandwidth of the spectrum analyzer was set to 300~Hz.

Fig.~\ref{Ampresults} shows the measurement results obtained with both
input beams amplitude squeezed.  $\hat S_{0}$, $\hat S_{1}$ and $\hat
S_{3}$ are all squeezed from 4.5~MHz to the limit of our measurement,
10~MHz.  $\hat S_{2}$ is anti-squeezed throughout the range of the
measurement.  Between 7.2~MHz and 9.6~MHz $\hat S_{0}$, $\hat S_{1}$
and $\hat S_{3}$ are all more than 3~dB below shot-noise.  The
squeezing of $\hat S_{0}$ and $\hat S_{3}$ is degraded at low
frequency due to phase noise on the second harmonic pump coupling into
the amplitude quadrature of the OPA outputs.  Since this noise is
correlated it cancels in the variance of $\hat S_{1}$.  The maximum
squeezing of $\hat S_{0}$ and $\hat S_{2}$ was 3.8~dB and 3.5~dB
respectively and was observed at 9.3~MHz.  The maximum squeezing of
$\hat S_{1}$ was 4.3~dB at 5.7~MHz.  The repetitive structure at 4, 5,
6, 7, 8 and 9~MHz was caused by electrical pick-up in our SHG
resonator emitted from a separate experiment operating in the
laboratory.

%
%
%
%

%
%
%
%

By locking both OPAs to amplification we obtained results similar to
those in Fig.~\ref{Ampresults} but with the input beams phase
squeezed.  In this case $ \hat S_{0}$, $\hat S_{1}$ and $\hat S_{3}$
were all anti-squeezed and $\hat S_{2}$ was squeezed throughout the
range of the measurement.  The optimum noise reduction of $\hat S_{2}$
was 2.8~dB below shot-noise and was observed at 4.8~MHz.
\begin{figure}[]
 \leftline{\includegraphics[width=8.4cm]{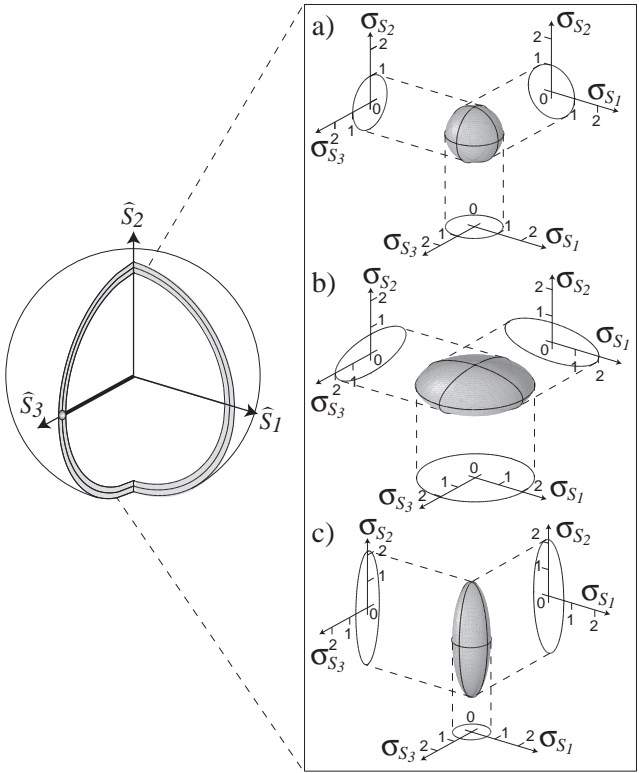}}
 \caption{Measured quantum polarization noise at 8.5~MHz mapped onto
 the Poincar\'{e} sphere.  a) coherent beam, b) beam from two phase
 squeezed inputs, c) beam from two amplitude squeezed inputs.  The
 surface of the ellipsoids defines the standard deviation of the noise
 normalized to shot-noise ($\sigma_{S_{i}} \!  = \!  \sqrt{V_{i}}$).}
 \label{Prettyresults}
\end{figure}

In Fig.~\ref{Prettyresults} we visualize our experimental data at
8.5~MHz assuming Gaussian noise statistics.  Given this assumption,
the standard deviation contour-surfaces shown here provide an accurate
representation of the states three dimensional noise distribution. 
The quantum polarization noise of a coherent state forms a sphere of
noise as portrayed in Fig.~\ref{Prettyresults}~a).  The noise volume
formed by our experimental results with two phase or two amplitude
squeezed beams respectively, was found to be a `pancake'-like
(Fig.~\ref{Prettyresults} b)) or `cigar'-like ellipsoid
(Fig.~\ref{Prettyresults} c)).

In all experiments prior to this work involving polarization squeezing
the squeezing was generated by combining a strong coherent beam with a
single weak amplitude squeezed beam \cite{GSYL87,HSSP99}.  Utilizing
only one squeezed beam it is not possible to simultaneously squeeze
any two of $\hat S_{1}$, $\hat S_{2}$, and $\hat S_{3}$ to quieter
than 3~dB below shot-noise ($V_{i} \!  + \!  V_{j} \!  \ge \! 
\expec{\hat n}$, with $i,j \in \{1,2,3; i \!  \not= \!  j\}$).  This
can be seen from the more general form of Eqs.~(\ref{var}) which
includes arbitrary phase angle $\theta$ between the beams.

To conclude, we present the first direct experimental characterization
of continuous variable quantum Stokes parameters.  We generate a new
quantum state with better than 3~dB squeezing of three Stokes
parameters ($\hat S_{0}$, $\hat S_{1}$, and $\hat S_{3}$)
simultaneously, a quality impossible to produce with only one squeezed
beam.  We represent our results on a quantum version of the
Poincar\'{e} sphere and experimentally demonstrate that in this space
a polarization squeezed beam generated from two amplitude squeezed
beams has a `cigar'-like quantum noise distribution; and one from two
phase squeezed beams has a `pancake'-like quantum noise distribution.

We would like to acknowledge the Alexander von Humboldt foundation for
support of Dr.~R.~Schnabel; the Australian Research Council for
financial support; Dr.~M.~B.~Gray and Dr.~B.~C.~Buchler for technical
assistance; and Dr.~T.~C.~Ralph for useful discussion.

\end{document}